\newcommand{\be}{\begin{equation}} 
\newcommand{\ee}{\end{equation}} 
\newcommand{\beeq}{\begin{eqnarray}} 
\newcommand{\eeeq}{\end{eqnarray}}

\def\Qdash{\overline{Q}}

\documentclass[12pt]{article}  
\usepackage{epsfig} 
\begin{document} 
\begin{flushright}                                                       
DESY  00-103\\                                                       
July 2000 \\                                                       
\end{flushright}                                                       
                                                       
\vspace*{1.5cm}                                                       
                                                       
\begin{center}                                                       
{\LARGE \bf Geometric scaling for the total $\gamma^{*}p$  cross section in  
 
the low $x$ region} \\              
              
                                                       
\vspace*{1cm}                                                       
{{\sc A.~M.~Sta\'sto}$^{(a,b)}$, {\sc K.~Golec-Biernat}$^{(b,c)}$ and
 {\sc J.~Kwieci\'nski}$^{(b)}$}  \\

\vspace*{0.5cm} 
    $^{(a)}$ \it INFN, Sezione di Firenze, Largo E. Fermi 2,\\
        \it 50125 Firenze, Italy\footnote{e-mail: stasto@fi.infn.it} \\

\vspace*{0.3cm}                 
$^{(b)}$ \it Department of Theoretical Physics,\\   
             \it H.~Niewodnicza\'nski Institute of Nuclear Physics,\\   
              \it ul. Radzikowskiego 152, 31-342 Krak\'ow,  
                  Poland\footnote{e-mail: jkwiecin@solaris.ifj.edu.pl}\\  
\vspace*{0.3cm} 

       $^{(c)}$ \it II Institut f\"ur Theoretische Physik,\\           
             \it Universit\"at Hamburg,\\                     
             \it Luruper Chaussee 149,                      
             \it 22761 Hamburg, Germany\footnote{e-mail: golec@mail.desy.de}\\

\end{center}                                                       
                                                       
\vspace*{1cm}                                                       
             
\begin{abstract}   
We observe that the saturation model of deep inelastic scattering, which   
successfully describes inclusive and diffractive data at small $x$,    
predicts a geometric scaling of the total   
$\gamma^{*}p$  cross section  in the region of small Bjorken variable $x$.  The   
geometric scaling in this case means that  the cross section is a function   
of only one dimensionless variable $\tau = Q^2 R_0^2(x)$,  where the function $R_0(x)$   
(called saturation radius) decreases with decreasing $x$.    
We show that the experimental data from   
HERA in the region $x<0.01$ confirm  the expectations of  this    
scaling over a very broad region of $Q^2$. We suggest that the geometric   
scaling is more general than the saturation  model.     
\end{abstract}  
 
\newpage 
    
It has recently been observed that the $ep$ deep inelastic scattering   
(DIS)     
data at low  $x$ \cite{HERA,FIXEDT} can be very economically described with     
the help of the saturation  model \cite{KGBW1}.  In this model the    
QCD dipole picture of the interaction between the    
transversely (T) or longitudinally (L) polarized     
virtual photon $\gamma^*$ (emitted by the incident electron)    
and the proton $p$ was adopted. In this case   
the total $\gamma^*p$ cross sections are given by  \cite{BJ,NIKO,MUELLER}   
\be    
\label{eq:sigma1}     
\sigma_{T,L}(x,Q^2)\, =\,      
\int d^2{\bf r} \int_0^1 dz\; |\Psi_{T,L}(r,z,Q^2)|^2\; \hat{\sigma}(r,x)\,,     
\ee   
where $\Psi_{T,L}$ is the {\bf wave function} for the splitting of      
the virtual photon into a $q\bar{q}$ pair (dipole),      
and $\hat{\sigma}$    
is the imaginary part of the forward scattering amplitude   
of the $q\bar{q}$ dipole on the proton, called   
the {\bf dipole cross section}, which    
describes  the interaction of the dipole with the proton.     
In addition, ${\bf{r}}$ is the transverse separation of the quarks in the   
$q\bar{q}$ pair,   and $z$ is the light-cone momentum fraction of the photon   
carried by the quark (or antiquark).   As usual, $-Q^2$ is the photon   
virtuality and $x$ is the Bjorken variable.   The cross section   
(\ref{eq:sigma1}) has a clear physical interpretation   in the proton rest   
frame in which the $q\bar{q}$ pair formation and a subsequent interaction with   
the proton are clearly separated in time.    
       
Let us recall that the standard DIS proton structure functions are related     
to $\sigma_{T,L}$ by     
\be     
\label{eq:sfrel}     
F_{T,L}(x,Q^2)\,=\,\frac{Q^2}{4\pi^2\alpha_{em}}\, \sigma_{T,L}(x,Q^2)\,,     
\ee     
and $F_2=F_T+F_L$.     
The wave function of the      
virtual photon is given by the following equations:      
\beeq     
\label{eq:wavet}     
|\Psi_{T}|^2 &=&     
\frac{3\,\alpha_{em}}{2\pi^2}\sum_f  e_f^2     
\left\{ [z^2+(1-z)^2] \Qdash_{f}^{2} K_{1}^{2}(\Qdash_f r)\right. \nonumber \\     
&& \left.+\, m_f^2\ K_{0}^{2}(\Qdash_f r)     
\right\}\,,     
\\ \nonumber     
\label{eq:wavel}     
|\Psi_{L}|^2 &=&     
\frac{3\,\alpha_{em}}{2\pi^2} \sum_f e_f^2     
\left\{     
4 Q^2 z^2(1-z)^2 K_0^2(\Qdash_f r)      
\right\}\,,     
\eeeq     
where the sum is performed over quarks with flavour $f$,      
charge $e_f$ and mass $m_f$, and     
\be     
\label{eq:qdash}     
\Qdash_{f}^{2}\, =\, z(1-z) Q^2+m_f^2\,.    
\ee   
The functions $K_{0,1}$ are the Bessel--Mc Donald functions.

The main assumption of the saturation model concerns      
the saturation property of the dipole cross section which is     
incorporated in the approach of ref. \cite{KGBW1} as below:     
\be    
\label{eq:sighat}     
\hat\sigma(x,r)\, =\, \sigma_0\; g\left(\frac{r}{R_0(x)}\right)\,.      
\ee   
The function $R_0(x)$ with the dimension of length, called saturation   
radius, decreases with   decreasing $x$, while the normalization $\sigma_0$ is   
independent of $x$.    When $\hat{r}\equiv r/R_0(x) \rightarrow \infty$      
the function $g(\hat{r})$ saturates to 1, so that      
$\hat \sigma(x,r) \rightarrow \sigma_0$.      
In the realization \cite{KGBW1} of the saturation model   
\be     
\label{eq:satmod}     
g(\hat{r})\,=\,1-\exp{(-\hat{r}^2/4)}\,.     
\ee     
The fact that the dipole cross section (\ref{eq:sighat})   
is limited   by the energy independent cross section $\sigma_0$ may be   
regarded   as a unitarity bound. This reflects the fact that the small $x$   
increase    of DIS structure functions generated by pure DGLAP or linear BFKL    
evolution  has to be tamed by unitarization effects     
\cite{GLR}-\cite{KOVCHEGOV1}.    
   
The characteristic feature of eq.~(\ref{eq:sighat})  is     
its ``geometric scaling'', i.e. $\hat{\sigma}(x,r)$   
only     
depends on the dimensionless ratio~$r/R_0(x)$, and its energy dependence     
is entirely driven by the saturation radius $R_0(x)$.     
The scaling property of (\ref{eq:sighat}) with $\sigma_0$ independent of $x$     
resembles geometric scaling of hadron-hadron scattering \cite{DDEUS}.     
In the latter case the relevant quantity is the scattering amplitude in the     
impact parameter representation $G(b^2,s)$, where $b$ is the impact parameter     
and $s$ is the CM energy squared.     
The geometric scaling in this case corresponds to the assumption that     
\be     
\label{eq:gscaling}     
G(b^2,s) \rightarrow G(\beta)\,,     
\ee     
where $\beta = {b^2/R^2(s) }$ with $R(s)$ corresponding to the     
interaction radius which increases with increasing energy.     
The analogy between scaling exhibited in hadron-hadron collisions     
and in deep inelastic scattering should not be taken too literally.     
For example, the two radii have different physical   
interpretation, and moreover, they show completely different energy   
dependence since  the saturation radius $R_0(x)$ decreases with increasing   
energy (for $x\simeq Q^2/s\rightarrow 0$).

The assumption about the scaling property of the dipole   
cross section (\ref{eq:sighat}) has  profound consequences for   
the measured $\gamma^*p$ cross section $\sigma_{\gamma^*   
p}=\sigma_{T}+\sigma_{L}$.  If we neglect the quark masses $m_f$ in the photon   
wave   functions  (\ref{eq:wavet}) we can   
rescale the dipole size $r \rightarrow r/R_0(x)$ in eq.~(\ref{eq:sigma1}) such   
that the integration variables are dimensionless. Thus,     
after the integration $\sigma_{\gamma^*p}$     
becomes a function of only one dimensionless variable $\tau=Q^2 R_0^2(x)$,    
instead of $x$ and $Q^2$ separately,       
\be     
\label{eq:scaling}     
\sigma_{\gamma* p}(x,Q^2)\;=\;\sigma_{\gamma* p}(\tau)\,,     
\ee     
where the scale for $\sigma_{\gamma* p}$ is provided by the   
dipole   cross section normalization $\sigma_0$. The non-zero light quark mass   
does not    lead to a significant breaking of the scaling 
(\ref{eq:scaling}).  Let us emphasize again that the new scaling for 
$\sigma_{\gamma^*p}$,  valid in the low $x$ region, is obtained due     
to the assumption that (\ref{eq:sighat}) depends on     
$r$ and $x$ through the dimensionless combination $r/R_0(x)$.      
Following the discussion in \cite{KGBW1} it is easy to show     
that with the realization (\ref{eq:satmod})      
we smoothly change the behaviour of  (\ref{eq:scaling}),     
\be     
\label{eq:approx}     
\sigma_{\gamma* p} \sim \sigma_0 ~~~~~~~~~\longrightarrow~~~~~~~~~~~~     
\sigma_{\gamma* p} \sim {\sigma_0/\tau}\,,     
\ee     
(modulo logarithmic  modifications in $\tau$) when      
$\tau$ changes from small to large values, respectively.     
The aim of this paper is to demonstrate that the DIS data do indeed     
approximately exhibit the geometric scaling (\ref{eq:scaling}) with the     
property (\ref{eq:approx}).

     
Let us discuss at first the parameterization of the saturation radius     
$R_0(x)$.      
In ref.~\cite{KGBW1} the saturation radius form was postulated as follows     
\be     
\label{eq:radold}     
R_0(x)\,=\,\frac{1}{Q_0}\,\left(\frac{x}{x_0}\right)^{\lambda/2}\,,     
\ee     
where $Q_0=1~\mbox{\rm GeV}$.  The parameters $x_0,~\lambda>0$ together     
with the dipole cross section normalization $\sigma_0$ were    
fitted to all     
inclusive DIS data with $x<0.01$.   
A very good description of data was obtained     
down to the region $Q^2<1 \; \rm GeV^2$.     
The saturation model was in particular able  to describe the transition    
from the region of   DIS to the region of low values of $Q^2$.   
Also DIS      
diffractive data were described without additional tuning of the parameters.     
For a recent related analyses see \cite{OTHER} and also \cite{DLF2}.   
     
Below we show that $R_0(x)$ can also be     
determined differently in a less model-dependent way.     
 To this aim let us observe that after suitable extension     
of the saturation model to the low $Q^2$ region including the photoproduction limit     
 $Q^2=0$,     
the  $x$ dependence of the saturation radius $R_0(x)$ can be correlated with     
the energy dependence of the total photoproduction cross section $\sigma_{\gamma p}$.     
In order to extend the saturation model to the region of low $Q^2$     
 we replace, following ref.~\cite{KGBW1}, $x$ by $\bar{x}$ defined as:     
\be     
\bar{x} = x \left( 1 + {4 m_f^2 \over Q^2} \right )     
\ee     
in the argument of $R_0(x)$ in eq.~(\ref{eq:sighat}) and keep $m_f \neq 0$.      
The variable $\bar{x}$ is related to the total energy      
$W$ by     
\be     
\label{eq:barxW}     
\bar{x} = {Q^2 + 4m_f^2 \over W^2}\,.     
\ee     
We note that the saturation model based on eqs.  
(\ref{eq:sigma1})-(\ref{eq:satmod})  can now be extended     
down to the region $Q^2 = 0$.     
The photoproduction cross section is given by equations 
(\ref{eq:sigma1},\ref{eq:wavet})    
with $Q^2=0$,     
$\bar{Q}_f^2=m_f^2$ and with $x$ replaced by $\bar{x} = 4m_f^2/W^2$.      
The dominant contribution  to the photoproduction cross section comes from the     
integration region $1/m_f^2 \gg r^2 \gg R_0^2(x)$  in the corresponding integral     
on the right hand side in eq.~(\ref{eq:sigma1}).     
In this region we can set $m_f^2 K_1^2(m_f r) \simeq 1/r^2$ and     
$\hat{\sigma}(x,r) \simeq  \sigma_0$.     
This gives the following relation between photoproduction cross section and the     
saturation radius      
\be     
\label{eq:photprod}     
\sigma_{\gamma p}(W) \,=\, \bar{\sigma_0} \ln \left( {1 \over     
m_f^2\,R_0^2(\bar{x})} \right)\,.     
\ee     
The parameter $\bar{\sigma_0}$ is related to the     
overall normalization of the dipole cross section ${\sigma_0}$     
by $\bar{\sigma_0}=(2\alpha_{em}/3\pi) {\sigma_0}$.      
From equation (\ref{eq:photprod}) we finally obtain the following     
prescription for the saturation radius      
\be     
\label{eq:satrad}     
R_0^2(\bar{x}) = {1 \over m_f^2}\, \exp\left(-{\sigma_{\gamma p} \over     
\bar{\sigma_0}}\right) \, . \ee     
For $\sigma_{\gamma p }$ we take the Donnachie--Landshoff parameterization     
\cite{DL} \be     
\sigma_{\gamma p} \,=\, a\,\, \bar{x}^{-0.08}\,,     
\label{eq:dl}     
\ee     
where we set $m_f = 140~\mbox{\rm MeV}$ (following \cite{KGBW1})  in     
eqs.~(\ref{eq:barxW}) and (\ref{eq:satrad}). Using results of the fit     
presented in \cite{DL} we find \be     
a \,=\, 68\, \mu b\, \left({4 m_f^2 \over 1 \rm GeV^2} \right)^{0.08}\,.     
\ee     
For $\bar{\sigma_0}$ we set $23~{\mu b}$ to obtain a good description of data.

  Let us now confront the implications of geometric scaling (\ref{eq:scaling})   
with experimental     
data on deep inelastic scattering at low $x$.     
 In Fig.~1 we show experimental data \cite{HERA} on the total cross section     
$\sigma_{\gamma^* p}$ plotted versus scaling variable $\tau = Q^2 R_0^2(x)$,      
with $R_0(x)$ obtained from Eq. (\ref{eq:satrad}).     
We include all available data for $x<0.01$ in the range of   
$Q^2$ values between $0.045 \; \rm GeV^2$ and $450 \; \rm GeV^2$.   
   
We see that to the data exhibit geometric scaling over a very    
broad   region of $Q^2$. We can also clearly see the change of shape     
of the dependence of $\sigma_{\gamma^* p}$ on $\tau$ from the approximate     
$1/\tau$ dependence at large $\tau$ to the less steep dependence  at small $\tau$.     
Thus the asymptotic relations (\ref{eq:approx}) are also to a good   
approximation   confirmed.      
The approximate asymptotic $1/\tau$ dependence reflects the fact that the cross section     
$\sigma_{\gamma^* p}$      
scales as $1/Q^2$ (modulo logarithmic corrections) and its energy      
dependence is governed by $1/R_0^2(x)$.     
Less steep dependence corresponds to the fact that at small values of $\tau$     
the total cross section grows weaker with energy than $1/R_0^2(x)$      
due to saturation of the dipole cross section, see eq. (\ref{eq:sighat}).    
    
We also found a symmetry between the regions of large and small $\tau$      
for the function $\sqrt{\tau} \sigma_{\gamma^* p}$,  which is illustrated in     
Fig.~2. For the asymptotic values of $\tau$ this is a manifestation of the     
relations (\ref{eq:approx}). It is remarkable that Fig.~2      
seems to indicate the presence of symmetry of $\sqrt{\tau}     
\sigma_{\gamma^* p}$  with respect to the transformation      
\be     
\tau~~~\longleftrightarrow~~~{1/\tau}     
\ee     
in the whole region of $\tau$.     
     
We have also tried the power law parameterization for the radius,      
\be     
R_0^2(x) \sim x^{\lambda}\,,     
\label{eq:powerlaw}     
\ee     
where $0.3< \lambda< 0.4$, in particular the original form proposed     
in \cite{KGBW1} (see eq. (\ref{eq:radold}) ),      
and found that the data also exhibit the     
geometric scaling with this choice of parameterization.     
The approximate $1/\tau$ dependence at large $\tau$ corresponds to the      
$x^{-\lambda}$ behaviour     
of the proton structure function $F_2$ at large $Q^2$.     

In the photoproduction case,       
the parameterization (\ref{eq:powerlaw})    
combined with relation (\ref{eq:photprod}) would   
correspond   to the logarithmic dependence on energy of the photoproduction   
cross section, i.e.      
\be     
\sigma_{\gamma p} \sim \ln(\bar{x})     
\label{eq:logdep}     
\ee     
We do therefore find that both prescriptions for $\sigma_{\gamma p}$      
eq.~(\ref{eq:dl}) and eq.~(\ref{eq:logdep}) give numerically similar results.     
     
In Fig.~3 we show contours  corresponding to different values of variable $\tau$ in     
$(x,Q^2)$ plane together with  experimental points for these values of $\tau$.     
Geometric scaling means that $\sigma_{\gamma^* p}$ is constant     
along each line.     
To be precise for each value of $\tau$ we plot experimental points    
within the bin $(\ln(\tau) \pm \delta)$      
with $\delta=0.1$. We see from this figure that for each $\tau$     
there are several experimental     
points for which $x$ varies as much as two orders of magnitude    
and $Q^2$ changes by a factor of $4$.     
Despite of that, all points along each line in Fig.~3     
are transformed to a narrow spread of points for a particular     
value of $\tau$ in Figs.~1 and 2, thus exhibiting geometric scaling.     
     
In order to show that the geometric scaling is confined only to the small $x$ region     
we show a similar  plot as  in Fig.~1,  but for the experimental data with     
$x>0.01$ \cite{HERA,FIXEDT}.      
It is evident that the geometric scaling is significantly violated in the     
region of large $x$, see Fig.~4.      
     
We would like to  emphasize that the (approximate) geometric scaling      
should predominantly be regarded as a remarkable regularity of DIS experimental data      
at low $x$.      
   
Although this scaling has been inspired by the saturation      
model of ref.~\cite{KGBW1},     
its significance goes presumably far beyond this model. In its essence   
the new scaling is a manifestation of the presence of an internal scale 
(saturation scale)     
characterizing  dense partonic systems, $Q_s(x) \sim 1/R_0(x)$.   
This scale emerges from a pioneering work  of  \cite{GLR}, which was   
subsequently analyzed and generalized in   
\cite{MUELLERS}- \cite{KOVCHEGOV1}.    
In the analysis  \cite{BARLEV}, and more recently in \cite{KOVCHEGOV},    
the scaling properties similar to those postulated in (\ref{eq:sighat})   
were found.   
An independent formulation  \cite{VENUGOPALAN}  of the small $x$ processes, 
gives the same overall picture with the saturation scale.  
At a deeper level,  the geometric scaling for  small-$x$   
processes may reflect self   
similarity or conformal   symmetry of the underlying dynamics. More   
detailed studies are under way, see  
\cite{BAL}-\cite{KOVCHEGOV1}.

To sum up we have shown that the experimental data on deep inelastic      
$ep$ scattering at low $x$ exhibit geometric scaling, i.e.    
the total cross section $\sigma_{\gamma^* p}(x,Q^2)$ is the   
function   of only one dimensionless variable $\tau = Q^2 R_0^2(x)$. This   
regularity was found   to hold over the very broad range of $Q^2$      
from $0.045~\mbox{\rm GeV}^2$ to $450~\mbox{\rm GeV}^2$.      
It would be interesting to understand  in detail a possible dynamical origin     
of  this simple regularity.       

\section*{Acknowledgments}     

We thank Jochen Bartels and Genya Levin for useful comments and Larry McLerran  
for useful correspondance.     
This research has been partially  supported by the EU Framework TMR programme,      
contract FMRX-CT98-0194, Deutsche Forschungsgemeinschaft     
and the Polish Commitee for Scientific Research     
grants Nos. KBN 2P03B 120 19, 2P03B 051 19.      
\newpage
\begin{figure}[tb]        
\begin{center}
\epsfig{file=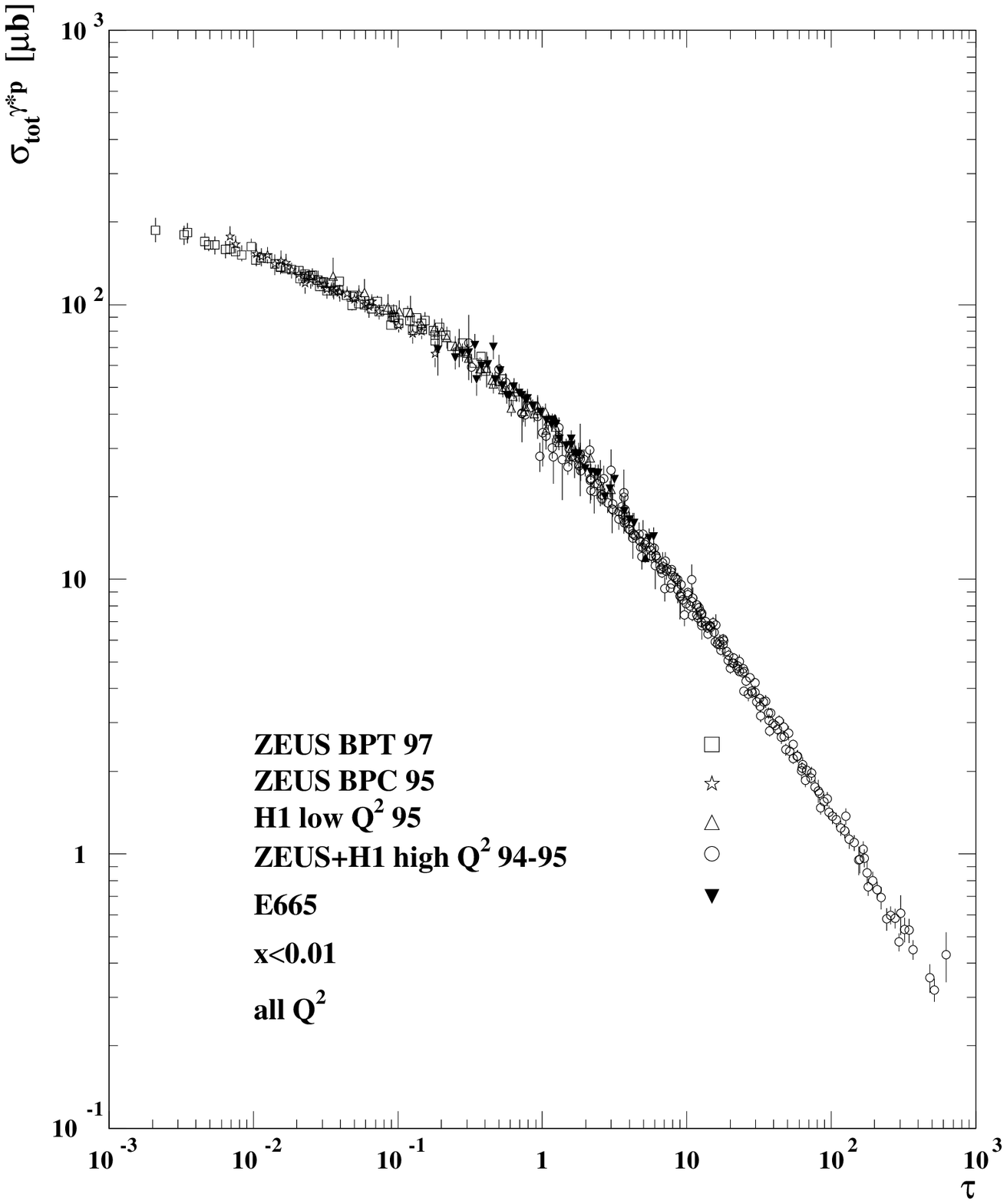,height=18cm,width=20cm}
\end{center}
\caption{Experimental data on $\sigma_{\gamma^* p }$      
from the region $x<0.01$ plotted versus      
the scaling variable $\tau=Q^2 R_0^2(x)$.}     
\label{fig1}     
\end{figure}     
\newpage
\newpage
\begin{figure}[tb]     
\centerline{\epsfig{file=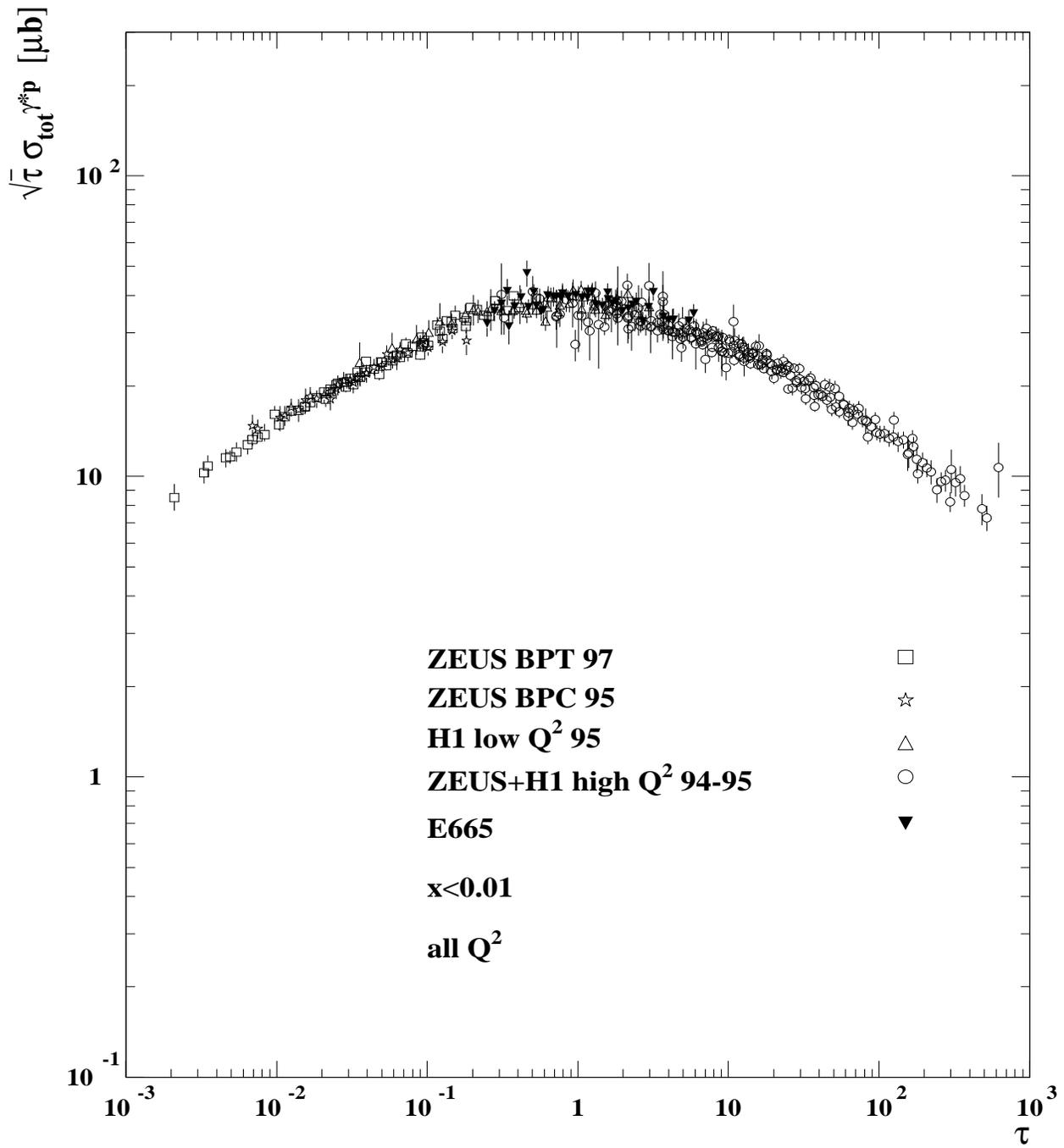,height=19cm,width=17cm}}     
\caption{ $\sqrt{\tau} \sigma_{\gamma^* p}$ plotted      
versus the scaling variable $\tau$}     
\label{fig2}     
\end{figure}    
\newpage
\begin{figure}[tb]     
\centerline{\epsfig{file=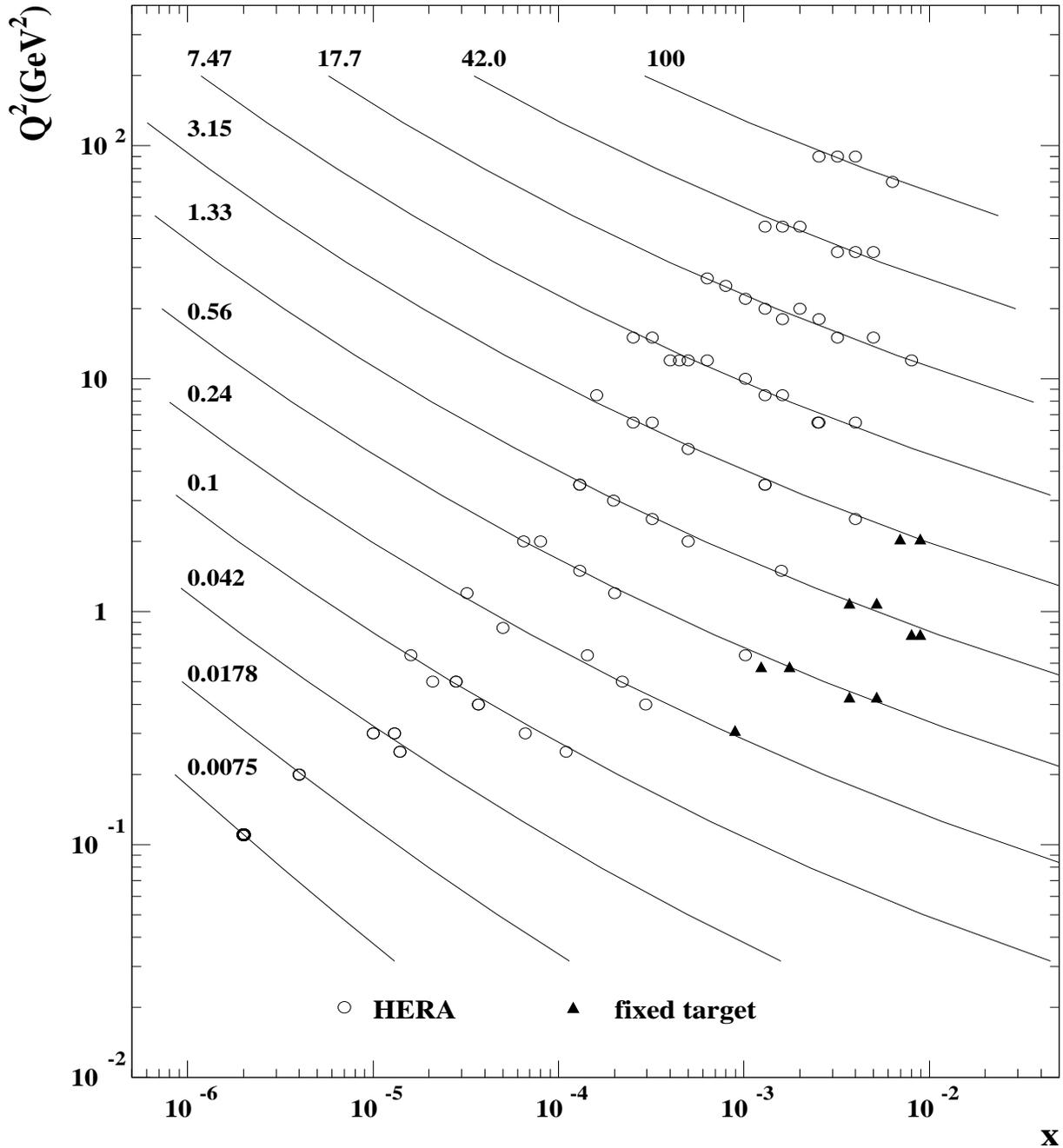,height=19cm,width=17cm}}     
\caption{The lines corresponding to different values of scaling     
variable $\tau$ (continues curves)     
in the $(x,Q^2)$-plane. The points correspond to available experimental data  located     
within the bins $\ln(\tau) \pm \delta$ ($\delta=0.1$) for each value of $\tau$. The numbers      
correspond to the value of $\tau$ for each curve.}     
\label{fig3}     
\end{figure}     
\newpage
\begin{figure}[tb]     
\centerline{\epsfig{file=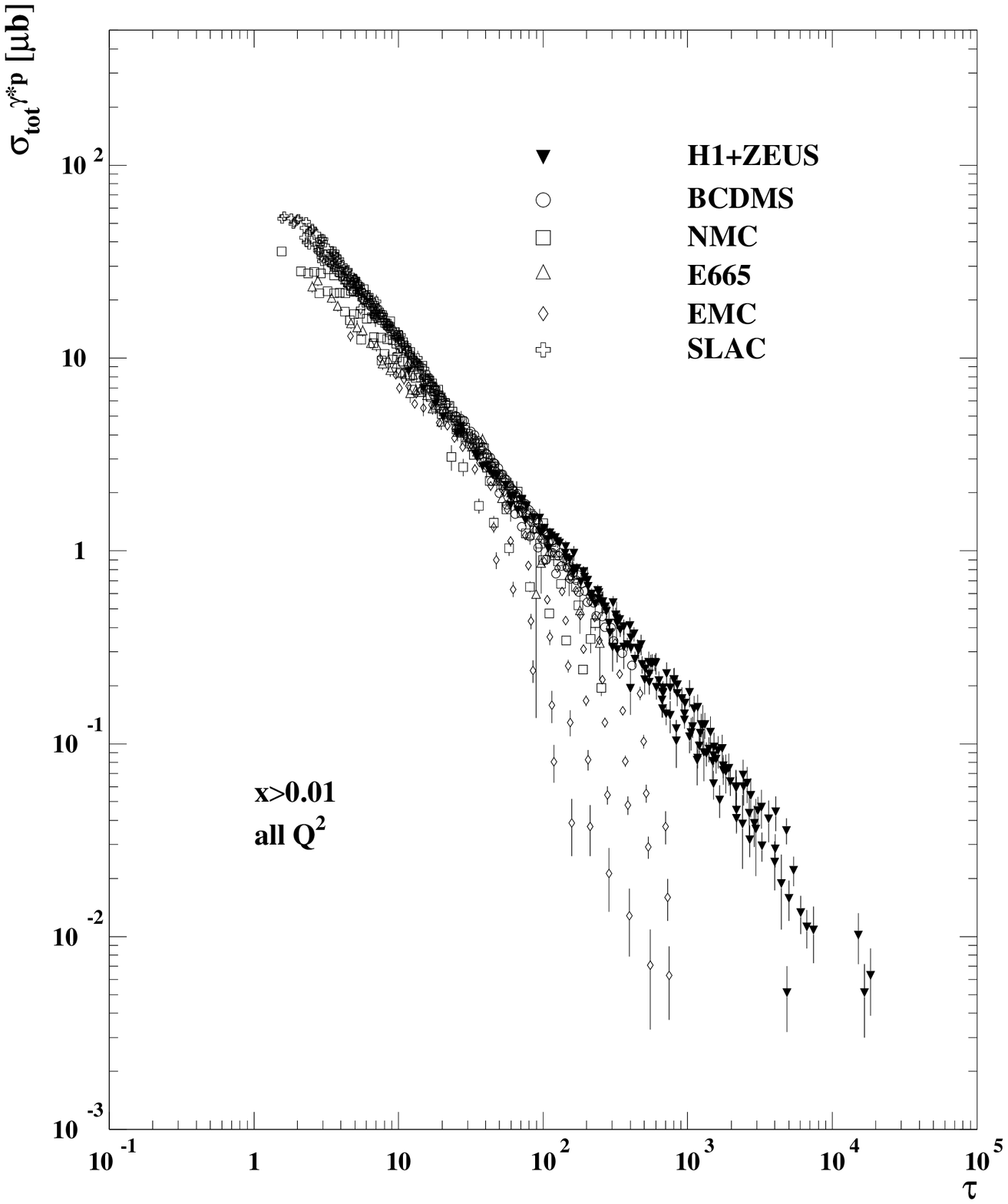,height=19cm,width=17cm}}     
\label{fig4}     
\caption{Experimental data on $\sigma_{\gamma^* p}$      
from the  region $x>0.01$ plotted versus      
the scaling variable $\tau=Q^2 R_0^2(x)$.}     
\end{figure}     
   

\newpage
\begin{thebibliography}{xx}   
\bibitem{HERA} H1 Collaboration (S. Aid et al.), {\em Nucl. Phys. } {\bf B 470} (1996) 3;   
ZEUS Collaboration (M. Derrick et al.), {\em Z. Phys. } {\bf C 72} (1996) 399;   
H1 Collaboration (C. Adloff et al.), {\em Nucl. Phys. } {\bf B 497} (1997) 3;     
ZEUS Collaboration (J. Breitweg et al.), {\em Phys. Lett. } {\bf B 407} (1997) 432;    
ZEUS Collaboration (J. Breitweg et al.), DESY-00-071, {\tt hep-ex/0005018}.    
\bibitem{FIXEDT} BCDMS collaboration (A.C.\ Benvenuti et al.), {\em   
Phys.\ Lett.} {\bf B 223} (1989) 485;    
NMC collaboration ( M.\ Arneodo et al.),    
{\em Phys. Lett. } {\bf B 364} (1995) 107;    
E665 collaboration ( M.R.\ Adams et al.),   
{\em Phys.\ Rev.} {\bf D 54} (1996) 3006;   
SLAC collaboration, {\em Phys. Lett.} {\bf B 282} (1992) 475;   
EMC  collaboration (Aubert et al.), {\em Nucl. Phys.} {\bf B 259} (1985) 189.   
 
 
 
\bibitem{KGBW1} K. Golec-Biernat and M. W\"usthoff,    
             {\em Phys. Rev.} \textbf{D59}  (1999) 014017;   
             {\em Phys. Rev.} \textbf{D60} (1999)  114023.    
%
 
\bibitem{BJ} J.D. Bjorken, J.B. Kogut and D.E. Sopper, {\em Phys. Rev.} {\bf 
D3} (1971) 1382. %
 
\bibitem{NIKO} N.N. Nikolaev and B.G. Zakharov, {\em Z. Phys.} {\bf C49}   
(1991) 607; {\em Z. Phys} {\bf C53} (1992) 331; {\em Z. Phys.} {\bf C64}   
(1994) 651; {\em JETP} {\bf 78} (1994) 598.   
 
%
\bibitem{MUELLER} A. H. Mueller, {\em Nucl.\ Phys.} {\bf B415} (1994) 373;   
A. H. Mueller and B. Patel, {\em Nucl.\ Phys.} {\bf B425} (1994) 471;   
A. H. Mueller, {\em Nucl. Phys.} {\bf B437} (1995) 107.  
   
\bibitem{GLR} L.V. Gribov, E.M. Levin and M.G. Ryskin, {\em Phys. Rep.}  
              {\bf 100}  (1983) 1.    
 
%
\bibitem{MUELLERS}  A.H. Mueller and  Jian-wei Qiu,   
{\em Nucl. Phys.} {\bf B268} (1986) 427 ;   
A.H. Mueller, {\em Nucl. Phys. } {\bf B335} (1990) 115; 
  Yu.A. Kovchegov, A.H. Mueller and S. Wallon, 
{\em Nucl. Phys.} {\bf B507} (1997) 367. 
A.H. Mueller,  
{\em Eur. Phys. J.} {\bf A1} (1998) 19;  
{\em Nucl.Phys.} {\bf  A654} (1999) 37c; 
{\em Nucl.Phys.} {\bf B558} (1999) 285. 
 
\bibitem{COLLINS} J. C. Collins and J. Kwieci\'nski, {\em Nucl. Phys. }  
{\bf B335} (1990) 89. 
 
\bibitem{BARTELS} J. Bartels, G.A. Schuler and  J. Bl\"umlein,  
{\em Z. Phys.} {\bf C50} (1991) 91;   
{\em Nucl. Phys. Proc. Suppl. } {\bf  18 C} (1991) 147; 
J. Bartels, {\em Phys. Lett.} {\bf B298} (1993) 204;   
{\em Z. Phys.} {\bf C60} (1993) 471; {\em Z. Phys.}  {\bf C62} (1994) 425; 
J. Bartels and M.  W\"usthoff,  {\em Z. Phys.} {\bf C66} (1995) 157; 
J. Bartels and C. Ewerz, {\em JHEP} {\bf 9909} (1999) 026.   
 
\bibitem{BARLEV} J. Bartels and G. Levin,  
{\em Nucl. Phys.} {\bf B387} (1992) 617. 
 
  
 
\bibitem{VENUGOPALAN} L. McLerran and  R. Venugopalan,  
{\em Phys. Rev.} {\bf D49} (1994) 2233, {\em Phys. Rev.} {\bf D49} (1994) 3352, 
{\em Phys. Rev.} {\bf D50} (1994) 2225; 
A. Kovner, L. McLerran and H. Weigert, 
{\em Phys.Rev.} {\bf D52} (1995) 6231, {\em Phys. Rev.} {\bf D52} (1995) 3809; 
R. Venugopalan, {\em Acta. Phys. Polon.} {\bf B30} (1999) 3731. 
 
 
   
\bibitem{BIALAS} A. Bia\l{}as and  R. Peschanski,   
     {\em Phys. Lett.} {\bf B355} (1995) 301; {\em Phys. Lett.} {\bf B378} (1996) 302; 
     {\em Phys. Lett.} {\bf B387} (1996) 405; 
     A. Bialas {\em Acta Phys. Polon.} {\bf B28} (1997) 1239; 
     A. Bialas and W. Czyz, {\em Acta Phys.Polon.} {\bf B29} (1998) 2095; 
     A. Bia\l{}as, H. Navelet and  R. Peschanski,    
     {\em Phys. Rev.} {\bf D57} (1998) 6585; 
      {\em Phys. Lett.} B427 (1998) 147; {\tt hep-ph/0009248}. 
      
\bibitem{SALAM} G.P. \ Salam, {\em Nucl.\ Phys.} {\bf B449} (1995) 589;   
{\em Nucl. Phys. } {\bf B461} (1996) 512;    
{\em Comput. Phys. Commun.} {\bf 105} (1997) 62;   
A.H. Mueller and  G.P. Salam, {\em Nucl.\ Phys.} {\bf B475} (1996) 293.   
 
 
\bibitem{LEVIN} E. Gotsman, E. M. Levin and  U. Maor,  
    {\em Nucl. Phys.} {\bf B464} (1996) 251;  
    {\em Nucl. Phys.} {\bf B493} (1997) 354; 
    {\em Phys. Lett.} {\bf B245} (1998) 369; 
    {\em Eur. Phys. J.} {\bf C5} (1998) 303;  
    E. Gotsman, E. M. Levin,  U. Maor and E. Naftali, 
   {\em Nucl. Phys.} {\bf B539} (1999) 535;    
    A. L. Ayala Filho, M. B. Gay Ducati and  E. M. Levin,   
   {\em Eur. Phys. J.} {\bf C8} (1999) 115; 
    E. Levin and U. Maor, {\tt hep-ph/0009217}. 
 
 
\bibitem{BAL} Ia. Balitsky, {\em Nucl.Phys. } {\bf B463}  (1996) 99. 
 
\bibitem{WEIGERT}  J. Jalilian-Marian, A. Kovner, L. McLerran  and  H. 
Weigert, {\em Phys. Rev.} {\bf D55} (1997) {5414}; 
J. Jalilian-Marian, A. Kovner and  H. 
Weigert, {\em Phys. Rev.} {\bf D59} (1999) {014014}; 
{\em Phys. Rev.} {\bf D59} (1999) {014015}; 
{\em  Phys. Rev.} D59 (1999) 034007; Erratum-ibid. D59 (1999) 099903; 
A. Kovner, J.Guilherme Milhano and  H. Weigert, 
OUTP-00-10P,NORDITA-2000-14-HE, {\tt hep-ph/0004014}; 
 H. Weigert, NORDITA-2000-34-HE, {\tt hep-ph/0004044}. 
 
 
\bibitem{KOVCHEGOV}  
Y.V. Kovchegov, {\em Phys. Rev.} {\bf D60}   
(1999) 034008;  {\em Phys. Rev.} {\bf D61} (2000) 074018; 
G. Levin and K. Tuchin, {\em Nucl. Phys.} {\bf B537} (2000) 833; 
M.A.  Braun, {\em  Eur. Phys. J.} {\bf C16} (2000) 337. 
 
\bibitem{KOVCHEGOV1} 
Y.V. Kovchegov and L. McLerran, 
{\em Phys. Rev.} {\bf D60} (1999) 054025; Erratum-ibid. {\bf D62} (2000) 
019901;  
Y.V. Kovchegov and G. Levin, {\em Nucl. Phys.} {\bf B577} (2000) 221; 
 
                      
%
\bibitem{DDEUS} J. Dias de Deus, {\em Nucl. Phys.}   
\textbf{B 59} (1973) 231;   A. J. Buras, J. Dias de Deus,  {\em Nucl. Phys. }   
\textbf{B 71} (1974) 481;  J. Dias de Deus, P. Kroll, {\em J. Phys. }   
\textbf{G 9} (1983) L81;  J. Dias de Deus, {\em Acta Phys. Polon. }   
\textbf{B 6} (1975) 613.   
 
\bibitem{OTHER}   W. Buchm\"uller, T. Gehrmann and A. Hebecker    
                 {\em Nucl. Phys.} \textbf{B537} (1999) 477;   
                  G. R. Forshaw, G. Kerley and G. Shaw,   
                 {\em Phys. Rev} \textbf{D60}  (1999) 074012, 
                 {\em Nucl. Phys.} {\bf A675} (2000) 80; 
                 E. Gotsman, E. Levin, U. Maor and E. Naftali,   
                 {\em Eur. Phys. J.} \textbf{C10} (1999) 689;                  
                 M. McDermott, L. Frankfurt, V. Guzey and M. Strikman,   
                 {\em Eur. Phys. J.} {\bf C16} (2000) 641;   
                  T. G. Cvetic, D. Schildknecht and  A. Shoshi,    
                 {\em Acta  Phys. Polon.} {\bf B 30} (1999) 3265;   
                 A. Capella, E. G. Ferreiro, A. B. Kaidalov and  
                 C. A. Salgado, {\tt  hep-ph/0006233}.   
 
\bibitem{DLF2} A. Donnachie and P.V. Landshoff, {\em Phys. Lett.} {\bf B437} 
(1998) 408.                     
%
\bibitem{DL}  A. Donnachie and P.V. Landshoff, {\em   
Phys. Lett. } \textbf{B296} (1992) 227.    
    
\end{thebibliography}
\end{document}